\documentclass[longauth]{aa}
\usepackage{graphicx}
\usepackage{txfonts}
\usepackage{xcolor}
\usepackage{hyperref}
\hypersetup{colorlinks, linkcolor=blue, citecolor=blue, urlcolor=blue}

\newcommand{\pivec}{\mbox{\boldmath $\pi$}}
\newcommand{\muvec}{\mbox{\boldmath $\mu$}}

\newcommand{\te}{t_{\rm E}}
\newcommand{\thetae}{\theta_{\rm E}}

\newcommand{\pie}{\pi_{\rm E}}
\newcommand{\pien}{\pi_{{\rm E},N}}
\newcommand{\piee}{\pi_{{\rm E},E}}
\newcommand{\dl}{D_{\rm L}}
\newcommand{\ds}{D_{\rm S}}
\newcommand{\hjdp}{{\rm HJD}^\prime}
\newcommand{\hjd}{{\rm HJD}}

\definecolor{brown}{rgb}{0.59, 0.29, 0.0}
\definecolor{darkgreen}{rgb}{0.0, 0.42, 0.24}
\definecolor{darkblue}{rgb}{0.01, 0.31, 0.59}
\definecolor{darkblue}{rgb}{0.0, 0.25, 0.42}
\definecolor{blue}{rgb}{0.0,0.0,1.0}
\definecolor{green}{rgb}{0.0,1.0,0.0}

\begin{document}

\title{KMT-2022-BLG-0475Lb and KMT-2022-BLG-1480Lb: Microlensing ice giants detected via non-caustic-crossing channel}
\titlerunning{KMT-2022-BLG-0475Lb and KMT-2022-BLG-1480Lb: Microlensing ice giants}

\author{
     Cheongho~Han\inst{01}
\and Chung-Uk~Lee\inst{02}
\and Ian~A.~Bond\inst{03}
\and Weicheng~Zang\inst{04,05}
\\
(Leading authors)\\
     Sun-Ju~Chung\inst{02, 04}
\and Michael~D.~Albrow\inst{06}
\and Andrew~Gould\inst{07,08}
\and Kyu-Ha~Hwang\inst{02}
\and Youn~Kil~Jung\inst{02,09}
\and Yoon-Hyun~Ryu\inst{02}
\and In-Gu~Shin\inst{04}
\and Yossi~Shvartzvald\inst{10}
\and Hongjing~Yang\inst{05}
\and Jennifer~C.~Yee\inst{04}
\and Sang-Mok~Cha\inst{02,11}
\and Doeon~Kim\inst{01}
\and Dong-Jin~Kim\inst{02}
\and Seung-Lee~Kim\inst{02}
\and Dong-Joo~Lee\inst{02}
\and Yongseok~Lee\inst{02,11}
\and Byeong-Gon~Park\inst{02}
\and Richard~W.~Pogge\inst{08}
\\
(The KMTNet collaboration)\\
     Shude~Mao \inst{05}
\and Wei~Zhu \inst{05}
\\
(Microlensing Astronomy Probe Collaboration)\\
     Fumio~Abe\inst{12}
\and Richard~Barry\inst{13}
\and David~P.~Bennett\inst{13,14}
\and Aparna~Bhattacharya\inst{13,14}
\and Hirosame~Fujii\inst{12}
\and Akihiko~Fukui\inst{15,16}
\and Ryusei~Hamada\inst{17}
\and Yuki~Hirao\inst{17}
\and Stela~Ishitani Silva\inst{13,18}
\and Yoshitaka~Itow\inst{12}
\and Rintaro~Kirikawa\inst{17}
\and Iona~Kondo\inst{17}
\and Naoki~Koshimoto\inst{19}
\and Yutaka~Matsubara\inst{12}
\and Shota~Miyazaki\inst{17}
\and Yasushi~Muraki\inst{12}
\and Greg~Olmschenk\inst{13}
\and Cl{\'e}ment~Ranc\inst{20}
\and Nicholas~J.~Rattenbury\inst{21}
\and Yuki~Satoh\inst{17}
\and Takahiro~Sumi\inst{17}
\and Daisuke~Suzuki\inst{17}
\and Taiga~Toda\inst{17}
\and Mio~Tomoyoshi\inst{17}
\and Paul~J.~Tristram\inst{22}
\and Aikaterini~Vandorou\inst{13,14}
\and Hibiki~Yama\inst{17}
\and Kansuke~Yamashita\inst{17}
\\
(The MOA Collaboration)\\
}

\institute{
      Department of Physics, Chungbuk National University, Cheongju 28644, Republic of Korea,                                                            
\and  Korea Astronomy and Space Science Institute, Daejon 34055, Republic of Korea                                                                       
\and  Institute of Natural and Mathematical Science, Massey University, Auckland 0745, New Zealand                                                       
\and  Center for Astrophysics $|$ Harvard \& Smithsonian, 60 Garden St., Cambridge, MA 02138, USA                                                        
\and  Department of Astronomy, Tsinghua University, Beijing 100084, China                                                                                
\and  University of Canterbury, Department of Physics and Astronomy, Private Bag 4800, Christchurch 8020, New Zealand                                    
\and  Max-Planck-Institute for Astronomy, K\"{o}nigstuhl 17, 69117 Heidelberg, Germany                                                                   
\and  Department of Astronomy, Ohio State University, 140 W. 18th Ave., Columbus, OH 43210, USA                                                          
\and  Korea University of Science and Technology, Korea, (UST), 217 Gajeong-ro, Yuseong-gu, Daejeon, 34113, Republic of Korea                            
\and  Department of Particle Physics and Astrophysics, Weizmann Institute of Science, Rehovot 76100, Israel                                              
\and  School of Space Research, Kyung Hee University, Yongin, Kyeonggi 17104, Republic of Korea                                                          
\and  Institute for Space-Earth Environmental Research, Nagoya University, Nagoya 464-8601, Japan                                                        
\and  Code 667, NASA Goddard Space Flight Center, Greenbelt, MD 20771, USA                                                                               
\and  Department of Astronomy, University of Maryland, College Park, MD 20742, USA                                                                       
\and  Komaba Institute for Science, The University of Tokyo, 3-8-1 Komaba, Meguro, Tokyo 153-8902, Japan                                                 
\and  Instituto de Astrof{\'i}sica de Canarias, V{\'i}a L{\'a}ctea s/n, E-38205 La Laguna, Tenerife, Spain                                               
\and  Department of Earth and Space Science, Graduate School of Science, Osaka University, Toyonaka, Osaka 560-0043, Japan                               
\and  Department of Physics, The Catholic University of America, Washington, DC 20064, USA                                                               
\and  Department of Astronomy, Graduate School of Science, The University of Tokyo, 7-3-1 Hongo, Bunkyo-ku, Tokyo 113-0033, Japan                        
\and  Sorbonne Universit\'e, CNRS, UMR 7095, Institut d'Astrophysique de Paris, 98 bis bd Arago, 75014 Paris, France                                     
\and  Department of Physics, University of Auckland, Private Bag 92019, Auckland, New Zealand                                                            
\and  University of Canterbury Mt.~John Observatory, P.O. Box 56, Lake Tekapo 8770, New Zealand                                                          
}


\abstract
{}
{
We investigate the microlensing data collected in the 2022 season from the high-cadence
microlensing surveys in order to find weak signals produced by planetary companions to lenses.
}
{
From these searches, we find that two lensing events KMT-2022-BLG-0475 and KMT-2022-BLG-1480 
exhibit weak short-term anomalies. From the detailed modeling of the lensing light curves, we 
identify that the anomalies are produced by planetary companions with a mass ratio to the 
primary of $q\sim 1.8\times 10^{-4}$ for KMT-2022-BLG-0475L and a ratio $q\sim 4.3\times 
10^{-4}$ for KMT-2022-BLG-1480L.
}
{
It is estimated that the host and planet masses and the projected planet-host separation are
$(M_{\rm h}/M_\odot, M_{\rm p}/M_{\rm U}, a_\perp/{\rm au}) = (0.43^{+0.35}_{-0.23}, 
1.73^{+1.42}_{-0.92}, 2.03^{+0.25}_{-0.38})$ for KMT-2022-BLG-0475L, and 
$(0.18^{+0.16}_{-0.09}, 1.82^{+1.60}_{-0.92}, 1.22^{+0.15}_{-0.14})$ 
for KMT-2022-BLG-1480L, where $M_{\rm U}$ denotes 
the mass of Uranus. Both planetary systems share common characteristics that the primaries of 
the lenses are early-mid M dwarfs lying in the Galactic bulge and the companions are ice giants
lying beyond the snow lines of the planetary systems.
}
{}


\maketitle

\section{Introduction}\label{sec:one}

The microlensing signal of a planet usually appears as a short-term anomaly on the smooth and 
symmetric lensing light curve generated by the host of the planet \citep{Mao1991, Gould1992b}. 
The signal arises when a source approaches the perturbation region formed around the caustic 
induced by the planet. Caustics represent the positions on the source plane at which the lensing 
magnification of a point source is infinite, and thus source crossings over the caustic result 
in strong signals with characteristic spike features.

The region of planetary deviations extends beyond caustics, and planetary signals can be produced 
without the caustic crossing of a source. Planetary signals produced via the non-caustic-crossing 
channel are weaker than those generated by caustic crossings, and the strength of the signal 
diminishes as the separation of the source from the caustic increases. Furthermore, these signals 
do not exhibit characteristic features such as the spikes produced by caustic crossings. Due to 
the combination of these weak and featureless characteristics, the planetary signals generated via 
the non-caustic channel are difficult to be noticed. If such signals are missed despite the fact
that they meet the criterion of detection, the statistical studies based on the incomplete planet 
sample would lead to erroneous results on the demographics of planets. In order to prevent this, 
the Korea Microlensing Telescope Network \citep[KMTNet:][]{Kim2016} group has regularly conducted 
systematic inspection of the data collected by the survey experiments in search of weak planetary 
signals and has published detected planets in a series of papers \citep{Zang2021a, Zang2021b, 
Zang2022, Zang2023, Hwang2022, Wang2022, Gould2022b, Jung2022, Jung2023, Han2022a, Han2022b, 
Han2022c, Han2022d, Han2022e, Han2023a, Shin2023}.

In this work, we present the analyses of the two microlensing events KMT-2022-BLG-0475 and
KMT-2022-BLG-1480, for which weak short-term anomalies were found from the systematic investigation 
of the data collected from the high-cadence microlensing surveys conducted in the 2022 season. We 
investigate the nature of the anomalies by carrying out detailed analyses of the light curves.

The organization of the paper for the presentation of the analyses and results are as follows. 
In Sect.~\ref{sec:two}, we describe the observations and data used in the analyses. In 
Sect.~\ref{sec:three}, we begin by explaining the parameters used in modeling the lensing 
light curves, and we then detail the analyses conducted for the individual events in the following 
subsections: Sect.~\ref{sec:three-one} for KMT-2022-BLG-0475 and in Sect.~\ref{sec:three-two} 
for KMT-2022-BLG-1480. In Sect.~\ref{sec:four}, we explain the procedure of constraining the 
source stars and estimating the angular Einstein radii of the events. In Sect.~\ref{sec:five}, 
we explain the procedure of the Bayesian analyses conducted to determine the physical lens 
parameters and present the estimated parameters. We summarize results and conclude in 
Sect.~\ref{sec:six}.

\section{Observations and data}\label{sec:two}

We inspected the microlensing data of the KMTNet survey collected from the observations
conducted in the 2022 season. The total number of KMTNet lensing events detected in the season
is 2803. For the individual events, we first fitted light curves with a single-lens single-source
(1L1S) model and then visually inspected residuals from the model. From this inspection, we
found that the lensing events KMT-2022-BLG-0475 and KMT-2022-BLG-1480 exhibited weak short-term
anomalies. We then cross-checked whether there were additional data from the surveys conducted
by other microlensing observation groups. We found that both events were additionally observed
by the Microlensing Observations in Astrophysics \citep[MOA:][]{Bond2001} group, who
referred to the events as MOA-2022-BLG-185 and MOA-2022-BLG-383, respectively. For
KMT-2022-BLG-1480, there were extra data acquired from the survey observations conducted by
the Microlensing Astronomy Probe (MAP) collaboration during the period from 2021 August to
2022 September, whose primary purpose was to verify short-term planetary signals found by the
KMTNet survey. In the analyses of the events, we used the combined data from the three survey
experiments.\footnote{ 
The Optical Gravitational Microlensing Experiment \citep[OGLE:][]{Udalski1994} is another major 
microlensing survey, although the two events analyzed in this work were not detected by the survey 
because the OGLE telescope was not operational in the first half of the 2022 season.  Besides 
these surveys dedicated to the microlensing program, lensing events are detected from other 
surveys such as the Zwicky Transient Facility (ZDF) survey \citep{Medford2023} and the Asteroid 
Terrestrial-impact Last Alert System (ATLAS) survey \citep{Tonry2018}, or observed using 
space-based instrument such as the Gaia survey \citep{Kruszynska2022, Luberto2022} and Hubble 
Space Telescope  \citep{Sahu2022}.
}

The observations of the events were carried out using the telescopes that are operated by the
individual survey groups. The three identical telescopes used by the KMTNet group have a 1.6~m
aperture equipped with a camera  yielding 4~deg$^2$ field of view, and they are distributed in 
the three continents of the Southern Hemisphere for the continuous coverage of lensing events. 
The sites of the individual KMTNet telescopes are the Cerro Tololo Interamerican Observatory in 
Chile (KMTC), the South African Astronomical Observatory in South Africa (KMTS), and the Siding 
Spring Observatory in Australia (KMTA). The MOA group utilizes the 1.3~m telescope at the 
Mt.~John Observatory in New Zealand, and the camera mounted on the telescope has a 1.2~deg$^2$ 
field of view. The MAP collaboration uses the 3.6~m Canada-France-Hawaii Telescope (CFHT) in 
Hawaii.

Observations by the KMTNet, MOA, MAP groups were done mainly in the $I$, customized MOA-$R$, 
and SDSS-$i$ bands, respectively.  A fraction of images taken by the KMTNet and MOA surveys 
were acquired in the $V$ band for the measurement of the source colors of the events.  Reduction 
of data and photometry of source stars were done using the pipelines of the individual survey 
groups. For the data used in the analyses, we readjusted the error bars estimated from the 
automated pipelines so that the error bars are consistent with the scatter of data and $\chi^2$ 
per degree of freedom for each data set becomes unity following the method described in 
\citet{Yee2012}.

\section{Light curve analyses}\label{sec:three}

The analyses of the lensing events were carried out by searching for lensing solutions 
specified by the sets of lensing parameters that best describe the observed light curves. 
The lensing parameters vary depending on the interpretation of an event. It is known that 
a short-term anomaly can be produced by two channels, in which the first is a binary-lens 
single-source (2L1S) channel with a low-mass companion to the lens, and the other is a 
single-lens binary-source (1L2S) channel with a faint companion to the source 
\citep{Gaudi1998}.

The basic lensing parameters used in common for the 2L1S and 1L2S models are $(t_0, u_0, \te, 
\rho)$.  The first two parameters represent the time of the closest source approach to the 
lens and the lens-source separation (impact parameter) scaled to the angular Einstein radius 
$\thetae$ at $t_0$, respectively. The third parameter denotes the event time scale, that is 
defined as the time for the source to transit $\thetae$. The last parameter is the normalized 
source radius, that is defined as the ratio of the angular source radius $\theta_*$ to $\thetae$. 
The normalized source radius is needed in modeling to describe the deformation of a lensing 
light curve caused by finite-source effects \citep{Bennett1996}.

In addition to the basic parameters, the 2L1S and 1L2S models require additional parameters 
for the description of the extra lens and source components. The extra parameters for the 
2L1S model are $(s, q, \alpha)$, where the first two parameters denote the projected separation 
scaled to $\thetae$ and the mass ratio between the lens components $M_1$ and $M_2$, and the 
last parameter denotes the source trajectory angle defined as the angle between the direction 
of the lens-source relative proper motion $\muvec$ and the $M_1$--$M_2$ binary axis. The extra 
parameters for the 1L2S model include $(t_{0,2}, u_{0,2}, \rho_2, q_F)$, which refer to the 
closest approach time, impact parameter, the normalized radius of the source companion $S_2$, 
and the flux ratio between the source companion and primary ($S_1$), respectively. See Table~2 
of \citet{Han2023b} for the summary of lensing parameters that are required to be included 
under various interpretations of the lens-system configuration.

For the individual events, we check whether higher-order effects improve the fits by conducting 
additional modeling. The considered higher-order effects are the microlens-parallax effect 
\citep{Gould1992a} and the lens-orbital effects \citep{Albrow2000}, which are caused by the 
orbital motion of Earth and lens, respectively. For the consideration of the microlens-lens 
parallax effects, we included two extra parameters $(\pien, \piee)$, which denote the north 
and east components of the microlens-parallax vector
\begin{equation}
\pivec_{\rm E} = \left( {\pi_{\rm rel}\over \thetae } \right) 
\left( { \muvec\over \mu} \right),
\label{eq1}
\end{equation}
respectively. Here $\pi_{\rm rel}=\pi_{\rm L}-\pi_{\rm S}={\rm au}(1/\dl - 1/\ds)$ denotes 
the relative lens-source parallax, while $\dl$ and $\ds$ denote the distances to the lens 
and source, respectively. The lens-orbital effects were incorporated into modeling by 
including two extra parameters $(ds/dt, d\alpha/dt)$, which represent the annual change 
rates of the binary-lens separation and the source trajectory angle, respectively.

We searched for the solutions of the lensing parameters as follows.  For the 2L1S modeling, we
found the binary-lens parameters $s$ and $q$ using a grid approach with multiple seed values 
of $\alpha$, and the other parameters were found using a downhill approach based on the MCMC
logic. For the local solutions identified from the $\Delta\chi^2$ map on the $s$--$q$ parameter 
plane, we then refined the individual solutions by allowing all parameters to vary. We adopted 
the grid approach to search for the binary parameters because it was known that the change of 
the lensing magnification is discontinuous due to the formation of caustics, and this makes it 
difficult to find a solution using a downhill approach with initial parameters $(s, q)$ lying 
away from the solution. In contrast, the magnification of a 1L2S event smoothly changes with 
the variation of the lensing parameters, and thus we searched for the 1L2S parameters using a 
downhill approach with initial values set by considering the magnitude and location of the 
anomaly features. In the following subsections, we describe the detailed procedure of modeling 
and present results found from the analyses of the individual events.

\begin{figure}[t]
\includegraphics[width=\columnwidth]{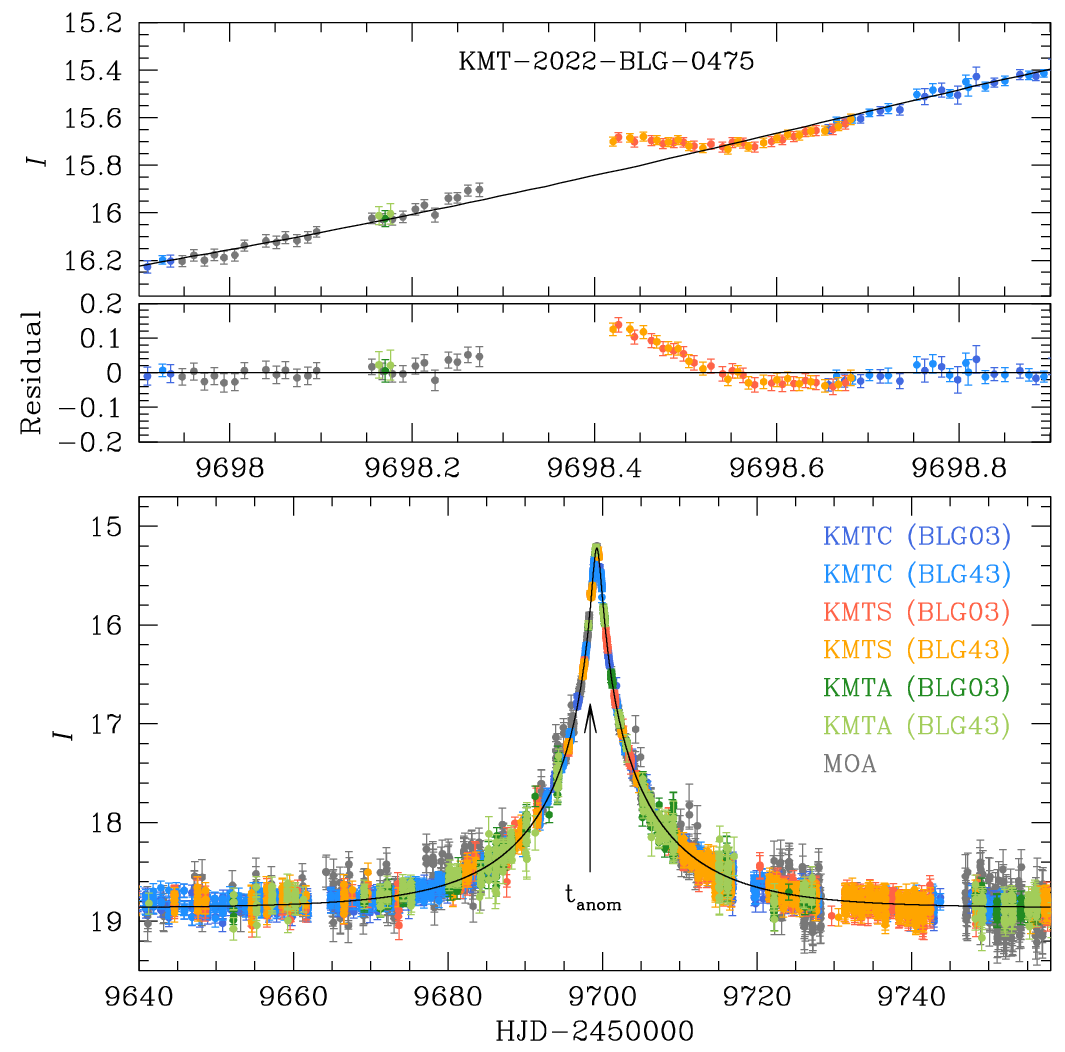}
\caption{
Light curve of KMT-2022-BLG-0475.  The lower panel shows the whole view and the upper panel 
shows the enlarged view of the region around the anomaly.  The arrow in the lower panel 
indicates the approximate time of the anomaly, $t_{\rm anom}$.  The curve drawn over the data 
points is the model curve of the 1L1S solution.  
The KMTC data set is used for the reference to align the other data sets.
}
\label{fig:one}
\end{figure}

\subsection{KMT-2022-BLG-0475}\label{sec:three-one}

The source of the lensing event KMT-2022-BLG-0475 lies at the equatorial coordinates $({\rm RA}, 
{\rm DEC})_{\rm J2000} =$ (18:05:20.56, -27:02:15.61), which correspond to the Galactic coordinates 
$(l, b) = (3^\circ\hskip-2pt .835, -2^\circ\hskip-2pt .804)$.  The KMTNet group first discovered 
the event  on 2022 April 19, which corresponds to the abridged Heliocentric Julian date $\hjdp\equiv 
\hjd-2450000=9688$, when the source was brighter than the baseline magnitude $I_{\rm base}=18.78$ 
by $\Delta I\sim 0.6$~mag.  Five days after the KMTNet discovery, the event was independently 
found by the MOA group, who designated the event as MOA-2022-BLG-185. Hereafter we use the KMTNet 
event notation following the convention of using the event ID reference of the first discovery 
survey. The event was in the overlapping region of the two KMTNet prime fields BLG03 and BLG43, 
toward which observations were conducted with a 0.5~hr cadence for each field and $\sim 0.25$~hr 
in combination.  The MOA observations were done with a similar cadence.

Figure~\ref{fig:one} shows the light curve of KMT-2022-BLG-0475 constructed from the combination 
of the KMTNet and MOA data. The anomaly occurred at around $t_{\rm anom}=9698.4$, which was $\sim 
0.85$~day before time of the peak. The zoom-in view of the region around the anomaly is shown the 
upper panel of Figure~\ref{fig:one}. The anomaly lasted for about 0.5~day, and the beginning part 
was covered by the MOA data while the second half of the anomaly was covered by the KMTS data. 
There is a gap between the MOA and KMTS data during $9698.30 \leq \hjdp \leq 9698.42$, and this 
gap corresponds to the night time at the KMTA site, which was clouded out except for the very 
beginning of the evening.

\begin{figure}[t]
\includegraphics[width=\columnwidth]{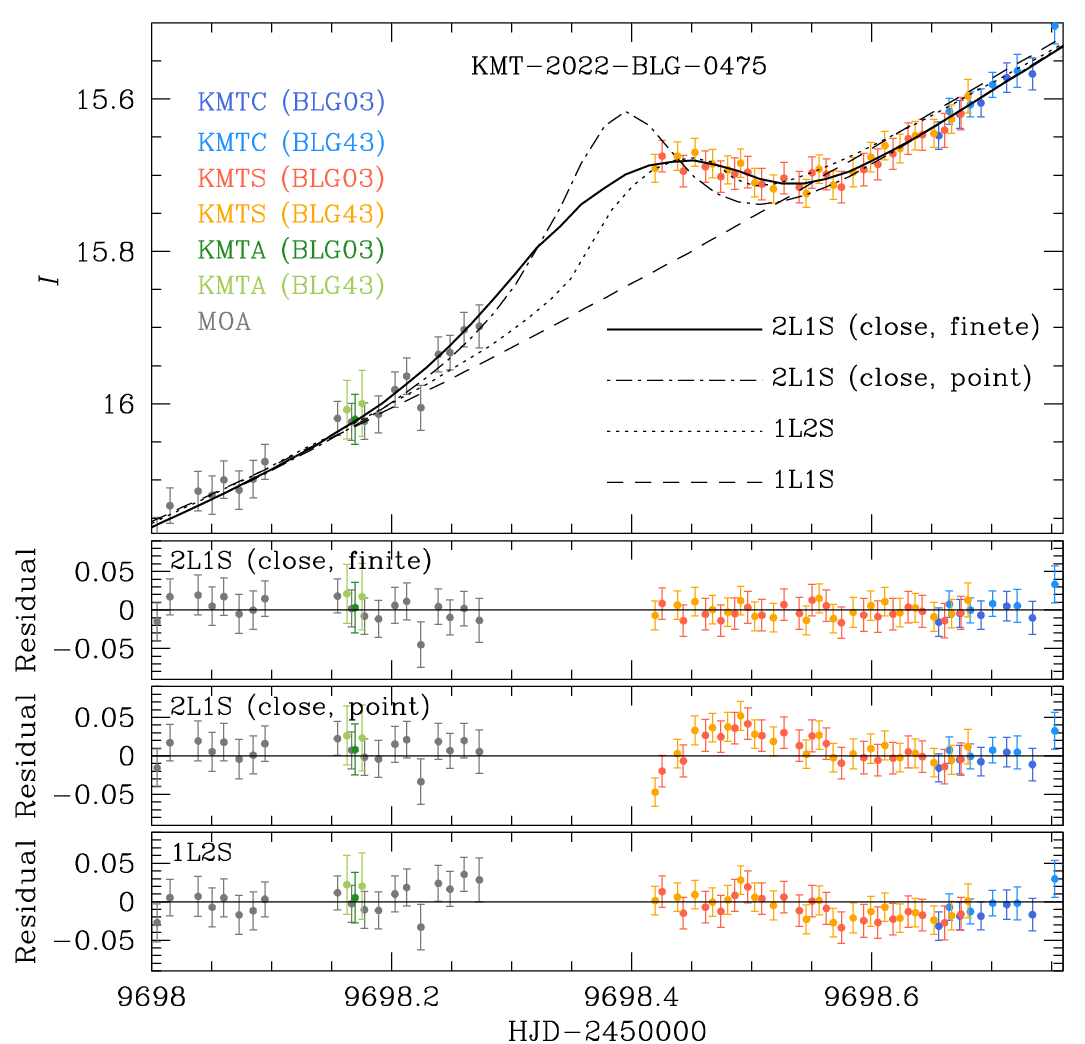}
\caption{
Zoom-in view around the anomaly in the lensing light curve of KMT-2022-BLG-0475. The lower three 
panels show the residuals from the finite- and point-source close 2L1S models and 1L2S model.
}
\label{fig:two}
\end{figure}

Figure~\ref{fig:two} shows the best-fit 2L1S and 1L2S models in the region around the anomaly. From 
the 2L1S modeling, we identified a pair of 2L1S solutions resulting from the close--wide degeneracy. 
In Table~\ref{table:one}, we present the lensing parameters of the two 2L1S and the 1L2S solutions 
together with the $\chi^2$ values of the fits and degrees of freedom (dof).  It was found that 
the severity of the degeneracy between the close and wide 2L1S solutions is moderate, with the 
close solution being preferred over the wide solution by $\Delta\chi^2=8.4$.  
For the best-fit solution, that is, the close 2L1S solution, we also list 
the flux values of the source $f_s$ and blend $f_b$, where the flux values are approximately 
scaled by the relation $I=18-2.5\log f$.

\begin{table*}[t]
\footnotesize
\caption{Model parameters of KMT-2022-BLG-0475\label{table:one}}
\begin{tabular}{l|cc|cc}
\hline\hline
\multicolumn{1}{c|}{Parameter}   &
\multicolumn{2}{c|}{2L1S}        &
\multicolumn{1}{c}{1L2S}         \\
\multicolumn{1}{c|}{ }           &
\multicolumn{1}{c}{Close}        &
\multicolumn{1}{c|}{Wide}        &
\multicolumn{1}{c}{}             \\
\hline
$\chi^2/{\rm dof}$            &  7903.4/7918             &  7911.8/7918             &   7930.7/7918            \\
$t_0$ ($\hjdp$)               &  $9699.258 \pm 0.002$    &  $9699.259 \pm 0.002$    &   $9699.276 \pm 0.003$   \\
$u_0$                         &  $0.035 \pm 0.001   $    &  $0.035 \pm 0.001   $    &   $0.035 \pm 0.001   $   \\
$\te$ (days)                  &  $16.84 \pm 0.13    $    &  $16.77 \pm 0.13    $    &   $17.05 \pm 0.13    $   \\
$s$                           &  $0.940 \pm 0.011   $    &  $1.135 \pm 0.012   $    &   --                     \\
$q$ (10$^{-4}$)               &  $1.76 \pm 0.26     $    &  $1.77 \pm 0.25     $    &   --                     \\
$\alpha$ (rad)                &  $5.692 \pm 0.004   $    &  $5.691 \pm 0.005   $    &   --                     \\
$\rho$ (10$^{-3}$)            &  $4.06 \pm  0.98    $    &  $3.62 \pm 1.09     $    &   --                     \\
$t_{0,2}$ ($\hjdp$)           &  --                      &  --                      &   $9698.425 \pm 0.012$   \\
$u_{0,2}$ (10$^{-2}$)         &  --                      &  --                      &   $-0.016 \pm 0.104  $   \\
$\rho_2$ (10$^{-3}$)          &  --                      &  --                      &   $3.51 \pm 0.69     $   \\
$q_F$ (10$^{-2}$)             &  --                      &  --                      &   $0.40 \pm 0.10     $   \\ 
$f_s$                         &  $0.4054 \pm 0.0003 $    &                          &     \\ 
$f_b$                         &  $-0.0215\pm 0.0010 $    &                          &     \\
\hline                                                 
\end{tabular}
\tablefoot{ ${\rm HJD}^\prime = {\rm HJD}- 2450000$.  }
\end{table*}

We find  that the anomaly in the lensing light curve of KMT-2022-BLG-0475 is best explained by a
planetary 2L1S model. The planet parameters are $(s, q)_{\rm close}\sim (0.94, 1.76\times 10^{-4})$ 
for the close solution and $(s, q)_{\rm wide}\sim (1.14, 1.77\times 10^{-4})$ for the wide solution. 
The estimated planet-to-host mass ratio is an order of magnitude smaller than the ratio between the 
Jupiter and the sun, $q\sim 10^{-3}$, indicating that the planet has a mass that is substantially 
smaller than a typical gas giant. Although the 1L2S model approximately describes the anomaly, it 
leaves residuals of 0.03~mag level in the beginning and ending parts of the anomaly, resulting in 
a poorer fit than the 2L1S model by $\Delta\chi^2=27.3$. It was found that the microlens-parallax 
parameters could not be measured because of the short time scale, $\te \sim 16.8$~days, of the event.

\begin{figure}[t]
\includegraphics[width=\columnwidth]{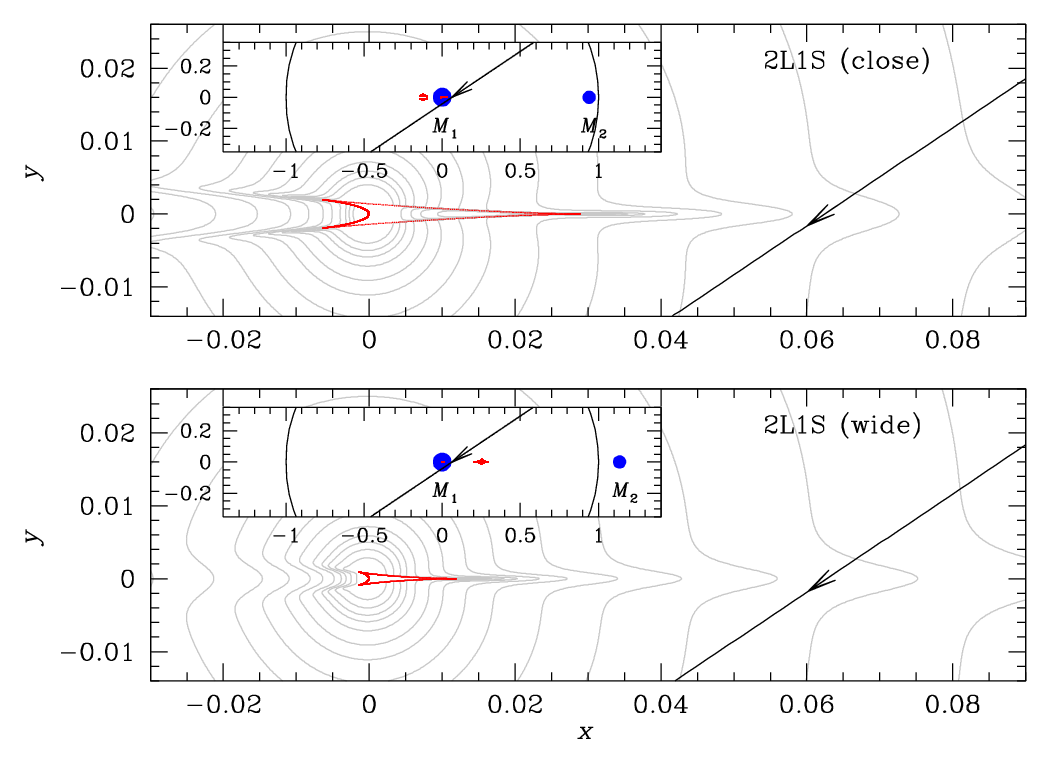}
\caption{
Lens-system configurations of the close (upper panel) and wide (lower panel) 2L1S solutions of 
KMT-2022-BLG-0475. In each panel, the red cuspy figures are caustics and the line with an arrow 
represents the source trajectory. The whole view of the lens system is shown in the inset, in 
which the small filled dots indicate the positions of the lens components and the solid circle 
represents the Einstein ring. The grey curves surrounding the caustic represent equi-magnification 
contours.
}
\label{fig:three}
\end{figure}

In the upper and lower panels of Figure~\ref{fig:three}, we present the lens-system configurations 
of the close and wide 2L1S solutions, respectively. In each panel, the inset shows the whole view 
of the lens system, and the main panel shows the enlarged view around the central caustic. A 
planetary companion induces two sets of caustics, with the "central" caustic indicating the one 
lying close to the primary lens, while the other caustic, lying away from the primary, is referred 
to as the as the "planetary" caustic.  The configuration shows that the anomaly was produced by 
the passage of the source through the deviation region formed in front of the protruding cusp of the 
central caustic.  We found that finite-source effects were detected despite the fact that the source 
did not cross the caustic.  In order to show the deformation of the anomaly pattern by finite-source 
effects, we plot the light curve and residual from the point-source model that has the same lensing 
parameters as those of the finite-source model except for $\rho$, in Figure~\ref{fig:two}.

It is known that the planet separations of the pair of degenerate solutions resulting from a
close--wide degeneracy follow the relation $\sqrt{s_{\rm close}\times s_{\rm wide}}=1.0$ 
\citep{Griest1998}.  For the close and wide solutions of KMT-2022-BLG-0475, this value is 
$\sqrt{s_{\rm close}\times s_{\rm wide}}=1.032$, which deviates from unity with a fractional 
discrepancy $(\sqrt{s_{\rm close}\times s_{\rm wide}}-1.0)/ 1.0\sim 3.2\%$.  We find that the 
relation between the two planet separations is better described by the \citet{Hwang2022} relation 
\begin{equation}
s^\dagger_\pm = \sqrt{s_{\rm in}\times s_{\rm out}} =
{\sqrt{u_{\rm anom}^2 +4}\pm u_{\rm anom} \over 2},
\label{eq2}
\end{equation}  
which was introduced to explain the relation between the planet separations $s_{\rm in}$ and 
$s_{\rm out}$ of the two solutions that are subject to the inner--outer degeneracy \citep{Gaudi1997}. 
Here $u_{\rm anom}^2=\tau^2_{\rm anom}+u_0^2$, $\tau_{\rm anom}=(t_{\rm anom}-t_0)/\te$, $t_{\rm anom}$ 
is the time of the anomaly, and the sign in the left and right sides of Eq.~(\ref{eq2}) is "$+$" for 
a major image perturbation and "$-$" for a minor-image perturbation. The terms "inner" and "outer" 
refer to the cases in which the source passes the inner and outer sides of the planetary caustic, 
respectively.  In the case of KMT-2022-BLG-0475 (and major-image perturbations in general), the 
close and wide solutions correspond to the outer and inner solutions, respectively.  From the 
measured planet separations of $s_{\rm in}=s_{\rm wide}=1.135$ and $s_{\rm out}=s_{\rm close}=
0.940$, we found that 
\begin{equation}
s^\dagger = (s_{\rm in}\times s_{\rm out})^{1/2} = 1.033. 
\label{eq3}
\end{equation}
From the lensing parameters $(t_0, t_{\rm anom}, \te, u_0)=$(9699.25, 9698.40, 16.8, 0.035), we found 
that $\tau =(t_{\rm anom}-t_0)/\te =0.050$, $u_{\rm anom}=0.061$, and
\begin{equation}
s^\dagger ={ \sqrt{u_{\rm anom}^2+4}+u_{\rm anom} \over 2} = 1.038. 
\label{eq4}
\end{equation}
Then, the fraction deviation of the $s^\dagger$ values estimated from Eqs.~(\ref{eq3}) and (\ref{eq4}) 
is $\Delta s^\dagger/s^\dagger=0.5\%$, which is 6.4 times smaller than the 3.2\% fractional discrepancy 
of the $\sqrt{s_{\rm close}\times s_{\rm wide}}=1$ relation.  
Although the $\sqrt{s_{\rm close}\times s_{\rm wide}}=1.0$ relation is approximately valid in the case 
of KMT-2022-BLG-0475, the deviation from the relation can be substantial especially when the planetary 
separation is very close to unity, and thus the relation in Eq.~(\ref{eq2}) helps to identify correct 
degenerate solutions.

\subsection{KMT-2022-BLG-1480}\label{sec:three-two}

The lensing event KMT-2022-BLG-1480 occurred on a source lying at $({\rm RA}, {\rm DEC})_{\rm J2000} 
= $ (17:58:54.96, -29:28:23.99), which correspond to $(l, b) = (1^\circ\hskip-2pt .015, -2^\circ\hskip-2pt 
.771)$.  The event was first found by the KMTNet group on 2022 July 11 ($\hjdp=9771$), when the source 
was brighter than the baseline magnitude, $I_{\rm base}=18.11$, by $\Delta I\sim 0.65$ mag. The source 
was in the KMTNet prime field BLG02, toward which observations were conducted with a 0.5~hr cadence. 
This field overlaps with the BLG42 field in most region, but the source was in the offset region that 
was not covered by the BLG42 field. The event was also observed by the MOA and MAP survey groups, who 
observed event with a 0.2~hr cadence and a 0.5--1.0~day cadence, respectively.

\begin{figure}[t]
\includegraphics[width=\columnwidth]{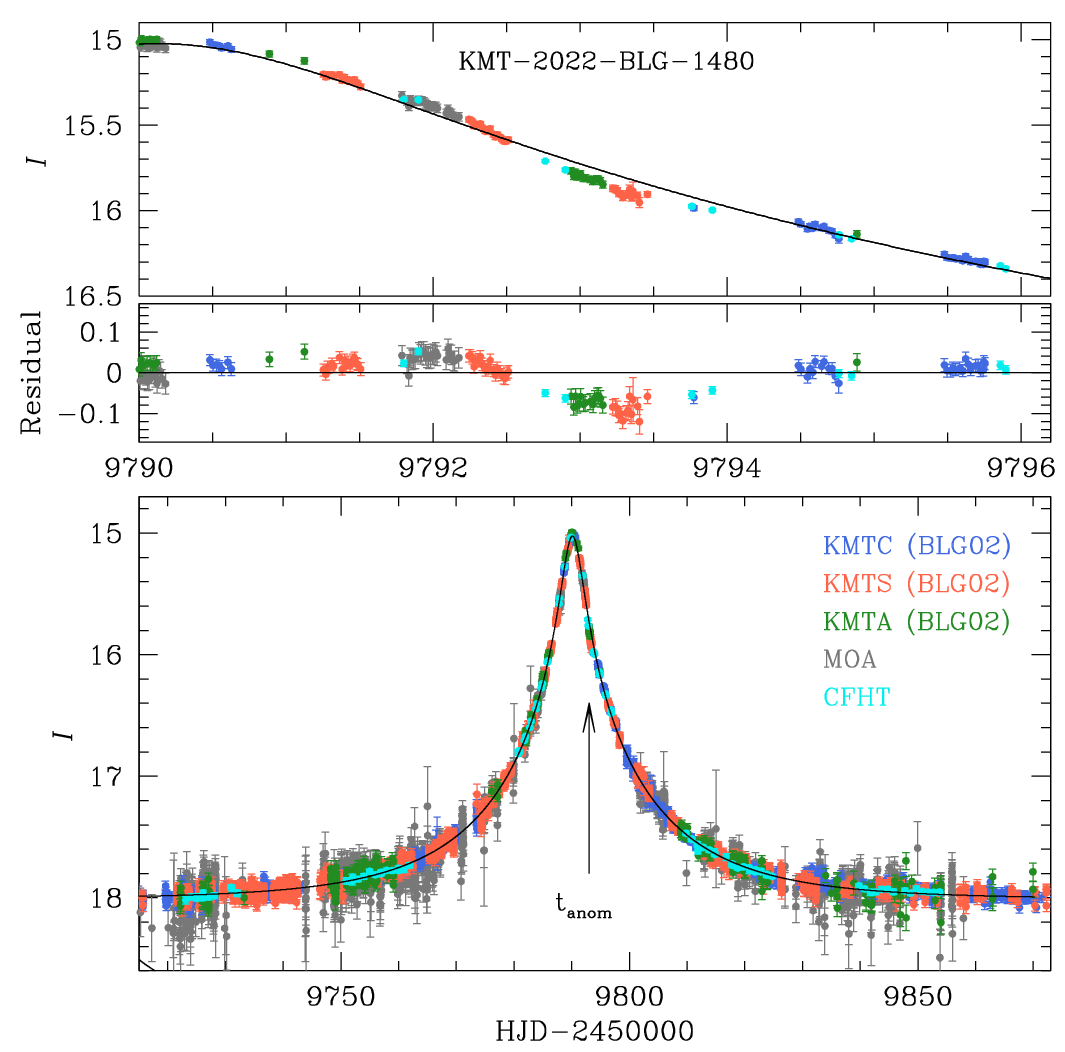}
\caption{
Light curve of the lensing event KMT-2022-BLG-1480.  Notations are same as those in Fig.~\ref{fig:one}.
}
\label{fig:four}
\end{figure}

In Figure~\ref{fig:four}, we present the light curve of KMT-2022-BLG-1480. We found that a weak 
anomaly occurred about 3 days after the peak centered at $t_{\rm anom}\sim 9793.2$. The anomaly 
is characterized by a negative deviation in most part of the anomaly and a slight positive deviation 
in the beginning part centered at $\hjdp\sim 9792.2$. The anomaly, that lasted about 2.7~days, was 
covered by multiple data sets from KMTS, KMTA, and CFHT. The sky at the KMTC site was clouded out 
for two consecutive nights from July 30 to August 1 ($9791 \leq \hjdp \leq 9793$), and thus the 
anomaly was not covered by the KMTC data.

\begin{figure}[t]
\includegraphics[width=\columnwidth]{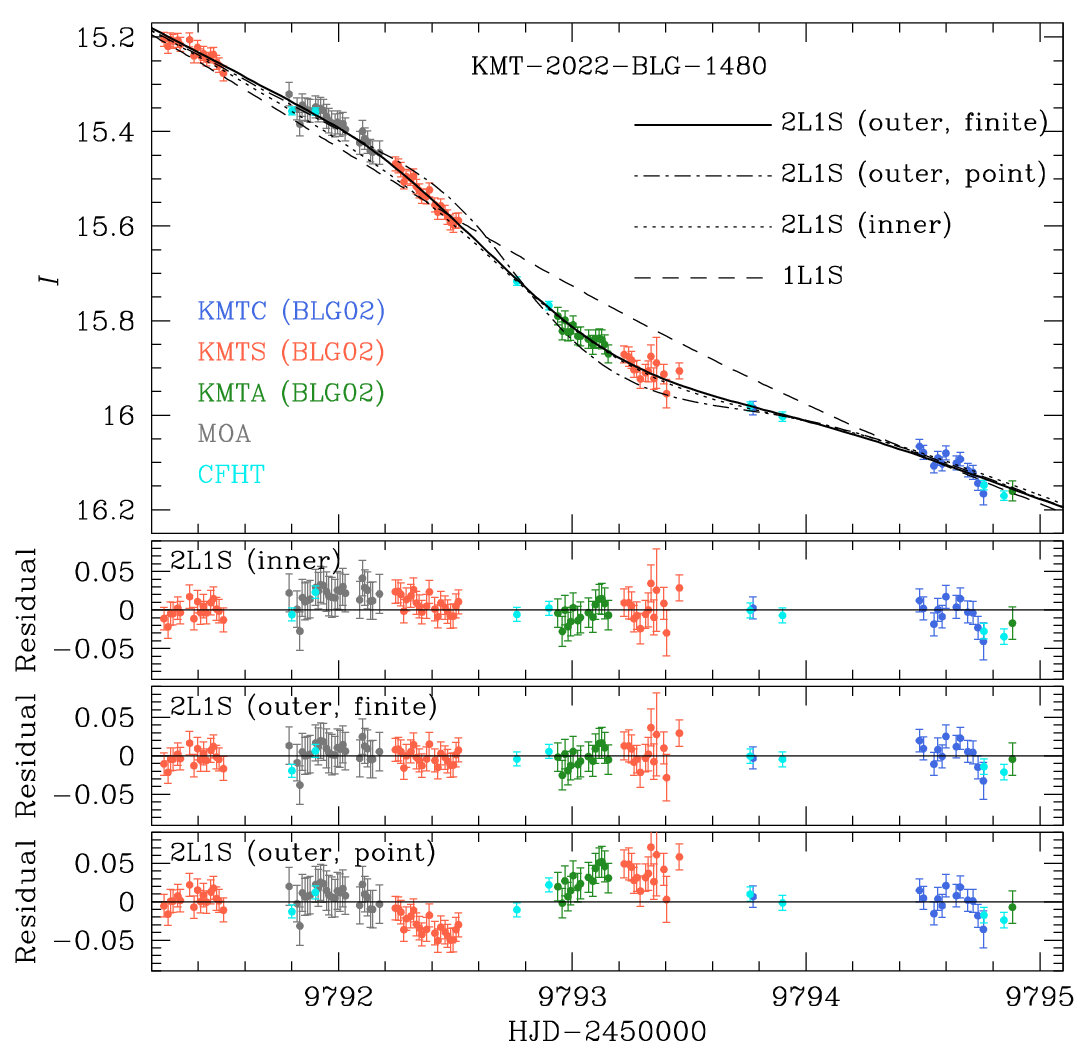}
\caption{
Enlarged view around the anomaly in the lensing light curve of KMT-2022-BLG-1480.  The lower 
three panels show the residuals form the inner 2L1S, finite- and point-source outer 2L1s models.
}
\label{fig:five}
\end{figure}

In Table~\ref{table:two}, we present the best-fit lensing parameters of the 2L1S solution. We did 
not conduct 1L2S modeling because a negative deviation cannot be explained with a 1L2S interpretation. 
We found a pair of 2L1S solutions, in which one solution has binary parameters $(s, q)\sim (0.83, 
4.9\times 10^{-4})$ and the other solution has parameters $(s, q)\sim (1.03, 4.7\times 10^{-4})$. 
Similar to the case of KMT-2022-BLG-0475L, the estimated mass ratio of order $10^{-4}$ is much 
smaller than the Jupiter/sun mass ratio. As we discuss below, the similarity between the model 
curves of the two 2L1S solutions is caused by the inner--outer degeneracy, and thus we refer to the 
solutions as "inner" and "outer" solutions, respectively. From the comparison of the inner and 
outer solutions obtained under the assumption of a rectilinear relative lens-source motion, it 
was found that the outer model yields a substantially better fit than the fit of the inner model 
by $\Delta\chi^2=63.9$, indicating that the degeneracy was resolved. In Figure~\ref{fig:five}, we 
present the model curves of the two solutions in the region around the anomaly. From the comparison 
of the models, it is found that the fit of the outer solution is better than the inner solution in 
the region around $\hjdp\sim 9792$, at which the anomaly exhibits slight positive deviations 
from the 1L1S model.

\begin{table*}[t]
\footnotesize
\caption{Model parameters of KMT-2022-BLG-1480\label{table:two}}
\begin{tabular}{l|c|ccc}
\hline\hline
\multicolumn{1}{c|}{Parameter}                 &
\multicolumn{1}{c|}{Inner}                     &
\multicolumn{3}{c}{Outer}                      \\
\multicolumn{1}{c|}{ }                         &
\multicolumn{1}{c|}{Standard}                  &
\multicolumn{1}{c}{Standard}                   &
\multicolumn{1}{c}{Higher-order ($u_0>0$)}      &
\multicolumn{1}{c}{Higher-order ($u_0<0$)}      \\
\hline
$\chi^2/{\rm dof}$            &   5724.8/5678              &   5660.9/5678             &   5654.0/5674            &   5655.3/5674            \\
$t_0$ (HJD$^\prime$)          &   $9790.133 \pm 0.004$     &   $9790.132 \pm 0.004$    &   $9790.134 \pm 0.004$   &   $ 9790.138 \pm 0.004$  \\
$u_0$                         &   $0.067 \pm 0.001   $     &   $0.069 \pm 0.001   $    &   $0.069 \pm 0.001   $   &   $   -0.069 \pm 0.001$  \\
$\te$ (days)                  &   $26.37 \pm 0.15    $     &   $26.09 \pm 0.15    $    &   $26.18 \pm 0.16    $   &   $   26.08 \pm 0.16  $  \\
$s$                           &   $0.826 \pm 0.004   $     &   $1.030 \pm 0.018   $    &   $1.017 \pm 0.014   $   &   $    1.011 \pm 0.015$  \\
$q$ ($10^{-4}$)               &   $4.87 \pm 0.36     $     &   $4.68 \pm 0.783    $    &   $4.30 \pm 0.69     $   &   $    4.18 \pm 0.70  $  \\
$\alpha$ (rad)                &   $0.516 \pm 0.004   $     &   $0.515 \pm 0.004   $    &   $0.525 \pm 0.006   $   &   $   -0.532 \pm 0.009$  \\
$\rho$ ($10^{-3}$)            &   $< 6               $     &   $14.18 \pm 4.02    $    &   $14.68 \pm 2.51    $   &   $   14.83 \pm 2.46  $  \\
$\pien$                       &   --                       &   --                      &   $0.42 \pm 0.29     $   &   $    0.27 \pm 0.31  $  \\
$\piee$                       &   --                       &   --                      &   $-0.02 \pm 0.05    $   &   $    0.04 \pm 0.05  $  \\
$ds/dt$ (yr$^{-1}$)           &   --                       &   --                      &   $0.60 \pm 0.40     $   &   $    1.14 \pm 0.44  $  \\
$d\alpha/dt$ (yr$^{-1}$)      &   --                       &   --                      &   $-0.57 \pm 0.49    $   &   $    1.59 \pm 0.73  $  \\
$f_s$                         &                            &  $0.9222 \pm 0.0009 $     &                          &                          \\ 
$f_b$                         &                            &  $0.0921 \pm 0.0016$      &                          &                          \\

\hline                                                 
\end{tabular}
\end{table*}

\begin{figure}[t]
\includegraphics[width=\columnwidth]{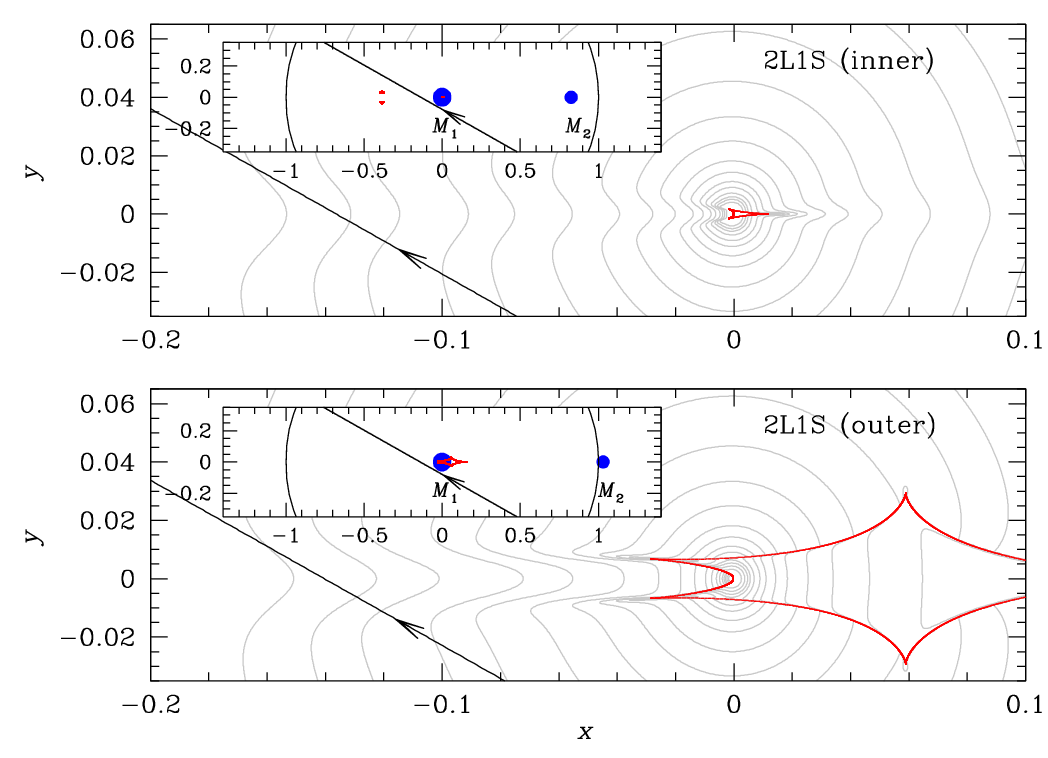}
\caption{
Lens-system configurations for the inner
and outer 2L1S solutions of KMT-2022-BLG-1480.
Notations are same as those in Fig.~\ref{fig:three}.
}
\label{fig:six}
\end{figure}

Figure~\ref{fig:six} shows the lens-system configurations of the inner and outer 2L1S solutions. 
Although the fit is worse, we present the configuration of the inner solution in order to find 
the origin of the fit difference between the two solutions. The configuration shows that the 
outer solution results in a resonant caustic, in which the central and planetary caustics merge 
and form a single caustic, while the central and planetary caustics are detached in the case of 
the inner solution. According to the interpretations of both solutions, the source passed the 
back-end side of central caustic without caustic crossings. The configuration of the outer 
solution results in strong cusps lying on the back-end side, and this caustic feature explains 
the slight positive deviation appearing in the beginning part of the anomaly around $\hjdp\sim 
9792.2$. Similar to the case of KMT-2022-BLG-0475, finite-source effects were detected although 
the source did not cross the caustic. We plot the point-source model in Figure~\ref{fig:five} 
for the comparison with the finite-source model.

We find that the relation in Eq.~(\ref{eq2}) is also applicable to the two local solutions of 
KMT-2022-BLG-1480. With $(s_{\rm in}, s_{\rm out})\sim (0.82, 1.03)$, the fractional deviation 
of the value $\sqrt{s_{\rm in}\times s_{\rm out}}$ from unity is $(\sqrt{s_{\rm in}\times 
s_{\rm out}}- 1.0)/1.0\sim 8\%$.  On the other hand, the fractional difference between $s^\dagger 
=\sqrt{s_{\rm in} \times s_{\rm out}}=0.919$ and $s^\dagger =[(u_{\rm anom}^2+4)^{1/2}-u_{\rm anom}]
/2=0.934$ is $\Delta s^\dagger /s^\dagger=1.6\%$, which is  5 times smaller than that of the 
$\sqrt{s_{\rm close} \times s_{\rm wide}}=1$ relation. This also indicates that the two local 
solutions result from the inner--outer degeneracy rather than the close--wide degeneracy.
We also checked the \citet{Hwang2021} relation between the planet-to-host mass ratio and the 
lensing parameters for the "dip-type" anomalies, 
\begin{equation}
q=\left( {\Delta t_{\rm dip} \over  4\te} \right)
{s\over |u_0| } |\sin\alpha|^3,
\label{eq5}
\end{equation}
where $\Delta t_{\rm dip}\sim 1.9$~day is the duration of the dip in the anomaly.  With the 
lensing parameters $(\te, u_0, s, \alpha)=(26, 0.069, 1.03, 59^\circ)$, we found that the mass 
ratio analytically estimated from Eq.~(\ref{eq5}) is $q\sim 5.9\times 10^{-4}$, which is close 
to the value $4.6\times 10^{-4}$ found from the modeling.

\begin{figure}[t]
\includegraphics[width=\columnwidth]{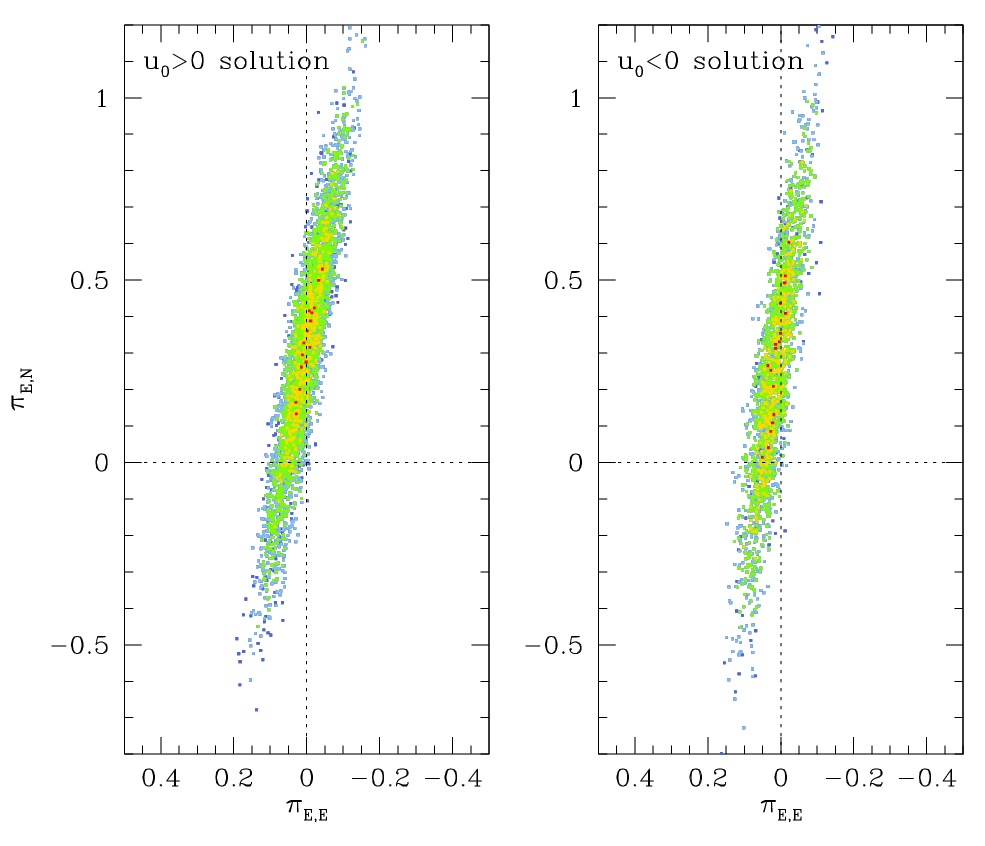}
\caption{
Maps of $\Delta\chi^2$ on the $(\piee, \pien)$ parameter plane 
for the higher-order $u_0 > 0$ and $u_0 < 0$ solutions of KMT-2022-BLG-1480.
Color coding is set to represent points with $\leq 1\sigma$ (red), $\leq 2\sigma$ (yellow), 
$\leq 3\sigma$ (green), $\leq 4\sigma$ (cyan), and $\leq 5\sigma$ (blue).
}
\label{fig:seven}
\end{figure}

\begin{table*}[t]
\footnotesize
\caption{Source properties, Einstein radii, and lens-source proper motions\label{table:three}}
\begin{tabular}{llll}
\hline\hline
\multicolumn{1}{c}{Quantity}                   &
\multicolumn{1}{c}{KMT-2022-BLG-0475}              &
\multicolumn{1}{c}{KMT-2022-BLG-1480}               \\
\hline
$(V-I, I)_{\rm S}$        &   $(2.016 \pm 0.005, 18.980 \pm 0.008)$   &   $(2.498 \pm 0.005, 18.088 \pm 0.006)$    \\
$(V-I, I)_{\rm RGC}$      &   $(2.108, 15.747)                    $   &   $(2.732, 16.199)                    $    \\
$(V-I, I)_{\rm 0,RGC}$    &   $(1.060, 14.332)                    $   &   $(1.060, 14.396)                    $    \\
$(V-I, I)_{\rm 0,S}$      &   $(0.968 \pm 0.005, 17.566 \pm 0.008)$   &   $(0.826 \pm 0.005, 16.284 \pm 0.006)$    \\
$\theta_*$ ($\mu$as)      &   $1.29 \pm 0.09                      $   &   $1.98 \pm 0.14                      $    \\
$\thetae$ (mas)           &   $0.32 \pm 0.02                      $   &   $0.13 \pm 0.01                      $    \\
$\mu$ (mas/yr)            &   $6.92 \pm 0.49                      $   &   $1.88 \pm 0.13                      $    \\
\hline                                                 
\end{tabular}
\end{table*}

We check whether the microlens-parallax vector $\pivec_{\rm E}=(\pien, \piee)$ can be measured 
by conducting extra modeling considering higher-order effects. We find that the inclusion of the 
higher-order effects improves the fit by $\Delta\chi^2 =6.9$ with respect to the model obtained 
under the rectilinear lens-source motion (standard model).
In Table~\ref{table:two}, we list the lensing parameters of the pair of higher-order models with 
$u_0 > 0$ and $u_0 < 0$, which result from the mirror symmetry of the source trajectory 
with respect to the binary-lens axis \citep{Skowron2011}. 
The $\Delta\chi^2$ maps of the models
on the $(\piee, \pien)$ 
parameter plane obtained from the higher-order modeling are shown in Figure~\ref{fig:seven}. 
It is found that the maps of the solutions results in a similar pattern of 
a classical 1-dimensional parallax ellipse, 
in which the east component $\piee$ is well 
constrained and the north component $\pien$ has a fairly big uncertainty.  \citet{Gould1994} 
pointed out that the constraints of the 1-dimensional parallax on the physical lens parameters 
are significant, and \citet{Han2016} indicated that the parallax constraint should be incorporated 
in the Bayesian analysis to estimate the physical lens parameters.  
We describe the detailed procedure of imposing the parallax constraint in the second paragraph of 
Sect.~\ref{sec:five}.  While the parallax parameters $(\pien, \piee)$ are constrained from the 
overall pattern of the light curve, the orbital parameters $(ds/dt, d\alpha/dt)$ are constrained 
from the anomaly induced by the lens companion.  For KMT-2022-BLG-1480, the orbital parameters are 
poorly constrained because the duration of the planet-induced anomaly is very short.

\begin{figure}[t]
\includegraphics[width=\columnwidth]{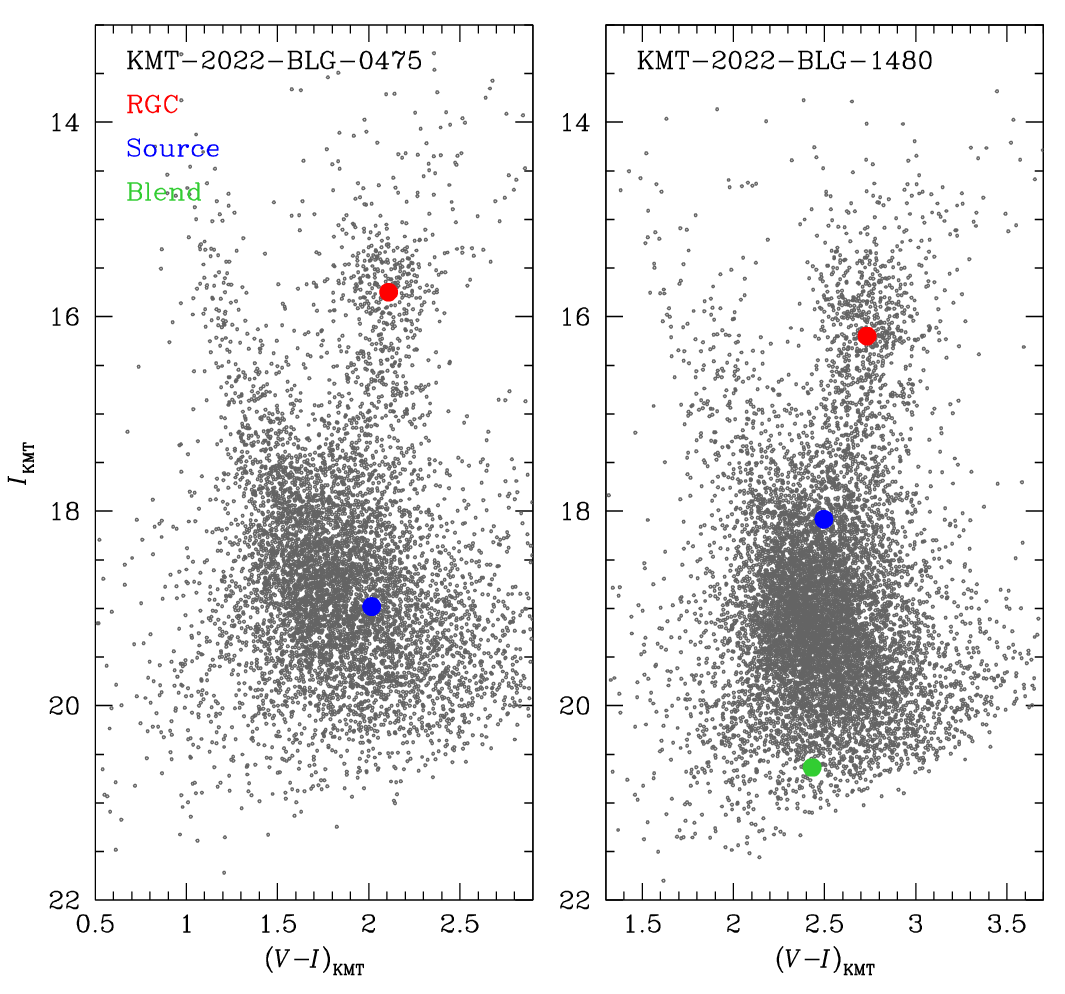}
\caption{
Source locations of the lensing events KMT-2022-BLG-0475 (left panel) and KMT-2022-BLG-1480 (right 
panel) with respect to the positions of the centroids red giant clump (RGC) in the instrumental 
color-magnitude diagrams of stars lying around the source stars of the individual events. For 
KMT-2022-BLG-1480, the position of the blend is additionally marked.
}
\label{fig:eight}
\end{figure}

\section{Source stars and angular Einstein radii}\label{sec:four}
   
In this section, we constrain the source stars of the events to estimate the angular Einstein 
radii. Despite the non-caustic-crossing nature of the planetary signals, finite-source effects 
were detected for both events, and thus the normalized source radii were measured. With the 
measured value of $\rho$, we estimated the angular Einstein radius using the relation
\begin{equation}
\thetae = {\theta_* \over \rho},
\label{eq6}
\end{equation}
where the angular radius of the source was deduced from the reddening- and extinction-corrected 
(de-reddened) color and magnitude.

The left and right panels of Figure~\ref{fig:eight} show the source locations in the instrumental 
color-magnitude diagrams (CMDs) of stars lying around the source stars of KMT-2022-BLG-0475 and 
KMT-2022-BLG-1480, respectively. 
For each event, the instrumental source color and magnitude, $(V-I, I)_{\rm S}$, were determined 
by estimating the flux values of the source $f_s$ and blend $f_b$ from the linear fit to the 
relation $F_{\rm obs} = A(t)f_s + f_b$, where the lensing magnification is obtained from the 
model.   We are able to constrain the blend for KMT-2022-BLG-1480 and mark its location on the 
CMD, but the blend flux of KMT-2022-BLG-0475 resulted in a slightly negative value, making it 
difficult to constrain the blend position.
By applying the method of \citet{Yoo2004}, we then estimated the de-reddened source 
color and magnitude, $(V-I, I)_{0,{\rm S}}$, using the centroid of the red giant clump (RGC), for 
which its de-reddened color and magnitude $(V-I, I)_{0,{\rm RGC}}$ were known \citep{Bensby2013, 
Nataf2013}, as a reference, that is,
\begin{equation}
(V-I, I)_{\rm 0,S} = (V-I, I)_{\rm 0,RGC} + [(V-I, I)_{\rm S} - (V-I, I)_{\rm RGC}].
\label{eq7}
\end{equation}
Here $(V-I, I)_{\rm RGC}$ denotes the instrumental color and magnitude of the RGC centroid, and 
thus the last term in the bracket represents the offset of the source from the RGC centroid in 
the CMD.

In Table~\ref{table:three}, we summarize the values of $(V-I, I)_{\rm S}$, $(V-I, I)_{\rm RGC}$, 
$(V-I, I)_{\rm 0,RGC}$, and $(V-I, I)_{\rm 0,S}$ for the individual events. From the estimated 
de-reddened colors and magnitudes, it was found that the source star of KMT-2022-BLG-0475 is an 
early K-type turnoff star, and that of KMT-2022-BLG-1480 is a late G-type subgiant.  We estimated 
the angular source radius by first converting the measured $V-I$ color into $V-K$ color using the 
\citet{Bessell1988} color-color relation, and then deducing $\theta_*$ from the \citet{Kervella2004} 
relation between $(V-K, V)$ and $\theta_*$. With the measured source radius, we then estimated the 
angular Einstein radius using the relation in Eq.~(\ref{eq6}) and the relative lens-source proper 
motion using the relation $\mu=\thetae/\te$.  The estimated $\thetae$ and $\mu$ values of the 
individual events are listed in Table~\ref{table:three}.  
We note that the uncertainties of the source colors and magnitudes presented in Table~\ref{table:three} 
are the values estimated from the model fitting, and those of $\theta_*$ and $\thetae$ are estimated by 
adding an additional 7\% error to consider the uncertain de-reddened RGC color of \citet{Bensby2013} 
and the uncertain position of the RGC centroid \citep{Gould2014}.

\section{Physical lens parameters}\label{sec:five}

We determined the physical parameters of the planetary systems using the lensing observables of
the individual events. For KMT-2022-BLG-0475, the measured observables are $\te$, and $\thetae$, 
which are respectively related to the mass and distance to the planetary system by
\begin{equation}
\te = {\thetae\over \mu}; \qquad
\thetae = (\kappa M \pi_{\rm rel})^{1/2},
\label{eq8}
\end{equation}
where $\kappa =4G/(c^2{\rm au})=8.14~{\rm mas}/M_\odot$. For KMT-2022-BLG-1480, we additionally 
measured the observable $\pie$, with which the physical parameters can be uniquely determined by
\begin{equation}
M= {\thetae \over \kappa \pie};\qquad
\dl = {{\rm au} \over \pie\thetae + \pi_{\rm S}}.
\label{eq9}
\end{equation}
We estimated the physical parameters by conducting Bayesian analyses because the observable $\pie$
was not measured for KMT-2022-BLG-0475, and the uncertainty of the north component of the parallax 
vector, $\pien$, was fairly big although the east component $\piee$ was relatively well constrained.

\begin{figure}[t]
\includegraphics[width=\columnwidth]{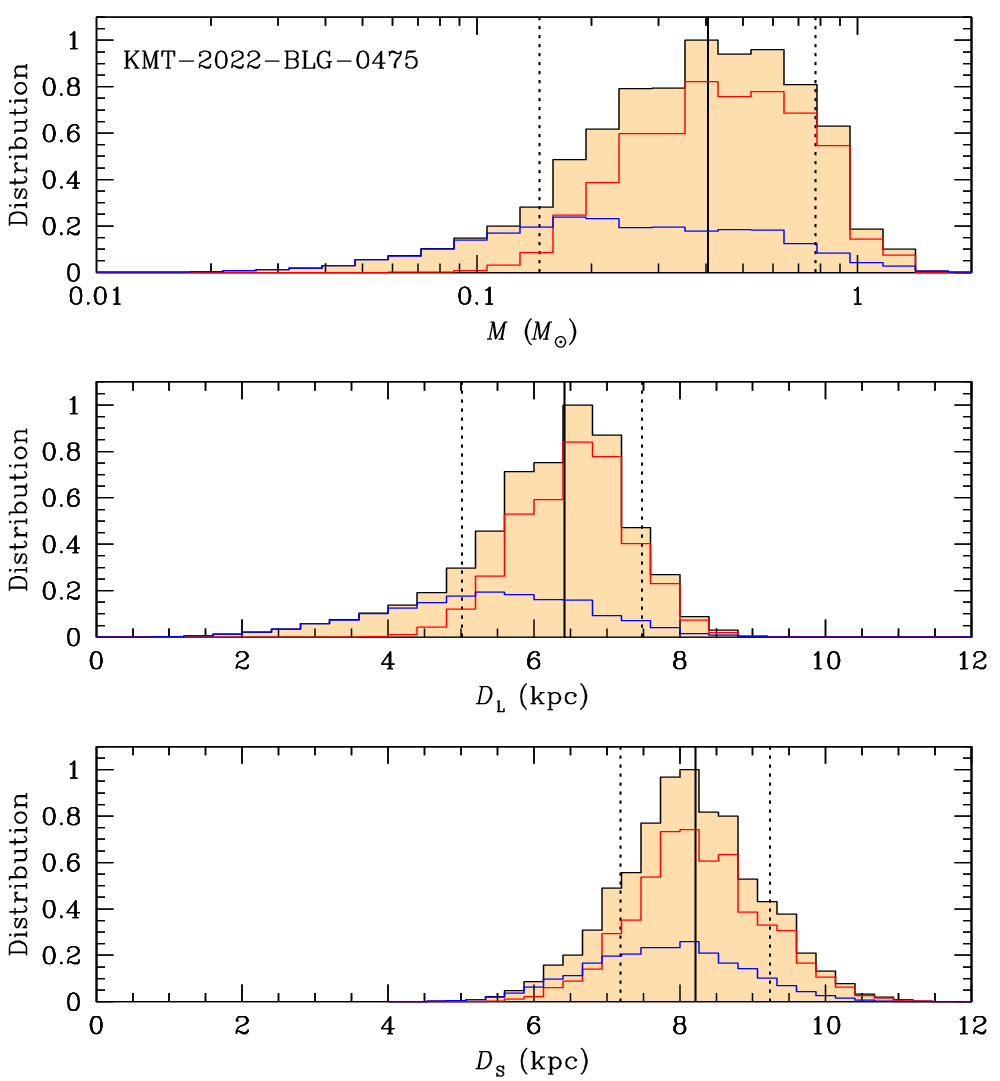}
\caption{
Bayesian posteriors of the lens mass and distances to the lens and source for the lens system 
KMT-2022-BLG-0475L. In each panel, the solid vertical line represents the median value, and 
the two dotted vertical lines indicate the uncertainty range of the posterior distribution.
The curves marked in blue and red represent the contributions by the disk and bulge lens 
populations, respectively.
}
\label{fig:nine}
\end{figure}

The Bayesian analysis was done by first generating artificial lensing events from a Monte Carlo
simulation, in which a Galactic model was used to assign the locations of the lens and source
and their relative proper motion, and a mass-function model was used to assign the lens mass.
We adopted the \citet{Jung2021} Galactic model and the \citet{Jung2018} mass function. With
the assigned values of $(M, \dl, \ds, \mu)$, we computed the lensing observables $(t_{{\rm E},i}, 
\theta_{{\rm E},i}, \pi_{{\rm E},i}$) of each simulated event using the relations in Eqs.~(\ref{eq1}) 
and (\ref{eq8}). 
Under the assumption that the physical parameters are independently and identically distributed,
we then constructed the Bayesian posteriors of $M$ and $\dl$ by imposing a weight
 $w_i = \exp(-\chi_i^2/2)$. Here the $\chi_i^2$ value for each event was computed by
\begin{equation}
\chi_i^2 = 
\left[ {t_{{\rm E},i}-\te \over \sigma(\te)}\right]^2 + 
\left[ {\theta_{{\rm E},i}-\thetae \over \sigma(\thetae)}\right]^2 + 
\sum_{j=1}^2 \sum_{k=1}^2 b_{j,k}
(\pi_{{\rm E},j,i}-\pi_{{\rm E},i})  (\pi_{{\rm E},k,i}-\pi_{{\rm E},i}),
\label{eq10}
\end{equation}
where $(\te, \thetae, \pie)$ represent the observed values of the lensing observables, $[\sigma(\te), 
\sigma(\thetae)]$ denote the measurement uncertainties of $\te$ and $\thetae$, respectively, $b_{j,k}$ 
denotes the inverse covariance matrix of $\pivec_{\rm E}$, and $(\pi_{{\rm E},1}, \pi_{{\rm E},2})_i 
= (\pien, \piee)_i$ denote the north and east components of the microlens-parallax vector of each 
simulated event, respectively.  We note that the last term in Eq.~(\ref{eq10}) was not included for 
KMT-2022-BLG-0475, for which the microlens-parallax was not measured.

In the case of the event KMT-2022-BLG-1480, for which the blending flux was measured, we additionally
imposed the blending constraint that was given by the fact that the lens could not be brighter than 
the blend. In order to impose the blending constraint, we computed the lens magnitude as
\begin{equation}
I_{\rm L} = M_{I,{\rm L}}+5\log \left( {\dl \over {\rm pc}} \right) -5 + A_{I,{\rm L}},
\label{eq11}
\end{equation}
where $M_{I,{\rm L}}$ represents the absolute $I$-band magnitude of a star corresponding to the lens 
mass, and $A_{I,{\rm L}}$ represents the $I$-band extinction to the lens. For the computation of 
$A_{I,{\rm L}}$, we modeled the extinction to the lens as
\begin{equation}
A_{I,{\rm L}} = A_{I,{\rm tot}} \left[ 1-\exp \left( -{ |z|\over h_{z,{\rm dust}}}  \right)\right],
\label{eq12}
\end{equation}
where $A_{I,{\rm tot}} = 1.53$ is the total $I$-band extinction toward the field, $h_{z,{\rm dust}} = 
100$~pc is the vertical scale height of dust, $z = \dl \sin b + z_0$, $b$ is the Galactic latitude, 
and $z_0=15$~pc is vertical position of the sun above the Galactic plane \citep{Siegert2019}.  It 
turned out that the blending constraint had little effect on the posteriors because the other constraints, 
that is, those from $(\te, \thetae, \pie)$, already predicted that the planet host are remote faint stars 
whose flux contribution to the blended flux is negligible.

\begin{figure}[t]
\includegraphics[width=\columnwidth]{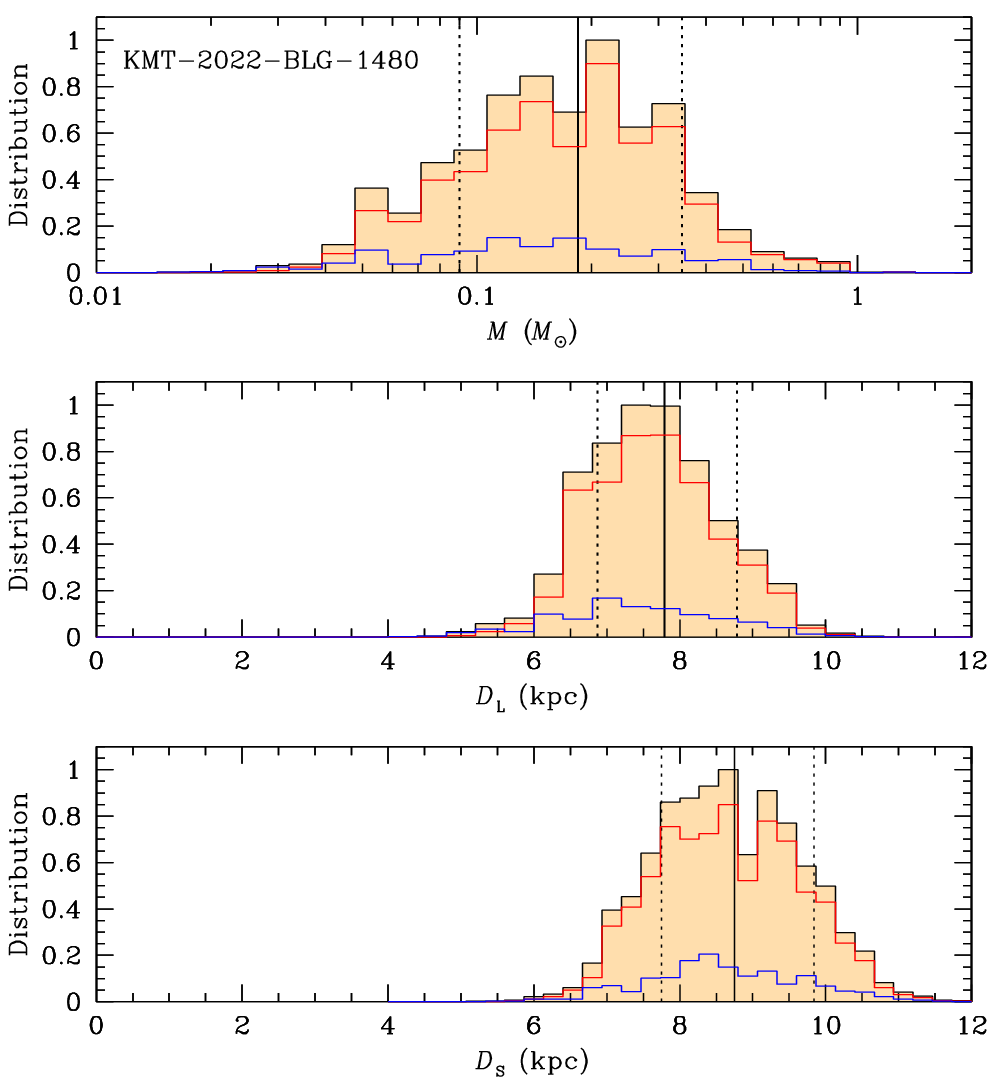}
\caption{
Bayesian posteriors of KMT-2022-BLG-1480.
Notations are same as those in Fig.~\ref{fig:nine}.
}
\label{fig:ten}
\end{figure}

Figures~\ref{fig:nine} and \ref{fig:ten} show the Bayesian posteriors of the lens mass and distances 
to the lens and source for KMT-2022-BLG-0475 and KMT-2022-BLG-1480, respectively. In 
Table~\ref{table:four}, we summarize the estimated parameters of the host mass, $M_{\rm h}$, planet 
mass, $M_{\rm p}$, distance to the planetary system, projected separation between the planet and host, 
$a_\perp=s\thetae \dl$, and the snow-line distances estimated as $a_{\rm snow}\sim 2.7(M/M_\odot)$
\citep{Kennedy2008}.
Here we estimated the representative values and uncertainties of the individual physical parameters 
as the median values and the 16\% and 84\% range of the Bayesian posteriors, respectively.

\begin{table}[t]
\footnotesize
\caption{Physical lens parameters\label{table:four}}
\begin{tabular*}{\columnwidth}{@{\extracolsep{\fill}}lcc}
\hline\hline
\multicolumn{1}{c}{Quantity}                   &
\multicolumn{1}{c}{KMT-2022-BLG-0475L}         &
\multicolumn{1}{c}{KMT-2022-BLG-1480L}         \\
\hline
$M_{\rm h}$ ($M_\odot$)       &  $0.43^{+0.35}_{-0.23}   $    &   $0.18^{+0.16}_{-0.09}   $    \\  [0.8ex]
$M_{\rm p}$ ($M_{\rm J}$)     &  $0.079^{+0.065}_{-0.042}$    &   $0.083^{+0.073}_{-0.042}$    \\  [0.8ex]
\hskip13pt ($M_{\rm U}$)      &  $1.73^{+1.42}_{-0.92}   $    &   $1.82^{+1.60}_{-0.92}   $    \\  [0.8ex]
$\dl$ (kpc)                   &  $6.58^{+0.82}_{-1.24}   $    &   $7.79^{+0.00}_{-0.92}   $    \\  [0.8ex]
$a_\perp$ (au)                &  $2.03^{+0.25}_{-0.38}   $    &   $1.22^{+0.15}_{-0.14}   $    \\  [0.8ex]
$a_{\rm snow}$ (au)           &  $1.16                   $    &   $0.49                   $    \\  [0.8ex]
\hline                                                 
\end{tabular*}
\tablefoot{  $M_{\rm J}$ and $M_{\rm U}$ denote the masses of Jupiter and Uranus, respectively.}
\end{table}

We find that the two planetary systems KMT-2022-BLG-0475L and KMT-2022-BLG-1480L are similar to each 
other in various aspects. According to the estimated physical lens parameters, the masses of 
KMT-2022-BLG-0475Lb and KMT-2022-BLG-1480Lb are $\sim 1.7$ and $\sim 1.8$ times the mass of Uranus in 
our solar system. The planets are separated in projection from their hosts by $\sim 2.0$~au and $\sim 
1.2$~au, respectively. The masses of planet hosts are $\sim 0.43~M_\odot$ and $\sim 0.18~M_\odot$, 
which correspond to the masses of early and mid-M dwarfs, respectively. Considering that the estimated 
separations are projected values and the snow-line distances of the planetary systems are $a_{\rm snow}
\sim 1.2$~au for KMT-2022-BLG-0475L and $\sim 0.5$~au for KMT-2022-BLG-1480L, the planets of both 
systems are ice giants lying well beyond the snow lines of the systems. The planetary systems lie at 
distances of $\sim 6.6$~kpc and $\sim 7.8$~kpc from the sun. 
The planetary systems are likely to be in the bulge with a probability 70\% for KMT-2022-BLG-0475L
and with a probability 83\% for KMT-2022-BLG-1480L.

\section{Summary and conclusion}\label{sec:six}

We analyzed the light curves of the microlensing events KMT-2022-BLG-0475 and KMT-2022-BLG-1480, 
for which weak short-term anomalies were found from the systematic investigation of the 2022 season 
data collected by high-cadence microlensing surveys. We tested various models that could produce the 
observed anomalies and found that the anomalies were generated by planetary companions to the lenses 
with a planet-to-host mass ratio $q\sim 1.8\times 10^{-4}$ for KMT-2022-BLG-0475L and a ratio $q\sim 
4.3\times 10^{-4}$ for KMT-2022-BLG-1480L. From the physical parameters estimated from the Bayesian 
analyses using the observables of the events, it was found that the planets KMT-2022-BLG-0475Lb and 
KMT-2022-BLG-1480Lb have masses $\sim 1.7$ and $\sim 1.8$ times the mass of Uranus in our solar system, 
respectively, and they lie well beyond the snow lines of their hosts of early and mid-M dwarfs, 
indicating that the planets are ice giants.

Ice giants around M dwarf stars are difficult to be detected by other surveys using the transit and 
radical-velocity (RV) methods not only because of the long orbital period of the planet but also 
because of the faintness of host stars. The number of low-mass planets increases with the increase 
of the observational cadence of microlensing surveys as shown in the histogram of detected 
microlensing planets as a function of the planet-to-host mass ratio presented in Fig.~1 of 
\citet{Han2022c}. Being able to complement the transit and RV surveys, high-cadence lensing surveys 
will play an important role in the construction of a more complete planet sample, and thus for better 
understanding the demographics of extrasolar planets.

The two events are also similar in that they have $\rho$ measurements (and therefore also
$\thetae$ measurements) despite the fact that the source does not cross any caustics.
\citet{Zhu2014} predicted that about half of KMT planets would not have caustic crossings, and
\citet{Jung2023} confirmed this for a statistical sample of 58 planetary events detected during the
2018--2019 period. However, \citet{Gould2022a} showed that about 1/3 of non-caustic-crossing events
nevertheless yield $\thetae$ measurements.

Measurements of $\thetae$ are important, not only because they improve the Bayesian estimates
(see Sect.~\ref{sec:five}), but also because they allow accurate prediction of when high-resolution
adaptive-optics (AO) imaging can resolve the lens separately from the source, which will then yield
mass measurements of both the host and the planet \citep{Gould2022a}. For KMT-2022-BLG-0475,
with proper motion $\mu=6.9$~mas/yr, the separation in 2030 (approximate first AO light on
30~m class telescopes), will be $\Delta\theta\sim 55$~mas, which should be adequate to resolve the
lens and source. 
Resolving the lens of this event would also be important to confirm the planetary interpretation of the 
event because it is difficult to completely rule out 
the 1L2S interpretation.
By contrast, for KMT-2022-BLG-1480, with $\mu=1.9$~mas/yr, the separation
will be only $\Delta\theta\sim 15$~mas, which almost certainly means that AO observations
should be delayed for many additional years. In particular, if the Bayesian mass and distance
estimates are approximately correct, then the expected contrast ratio between the source and lens
is $\Delta K\sim 7$~mag, which will likely require separations of at least 4 FWHM, that is,
55~mas even on the 39~m European Extremely Large Telescope. Hence, the contrast between the
two planets presented in this paper underlines the importance of $\thetae$ measurements.

\begin{acknowledgements}
Work by C.H. was supported by the grants of National Research Foundation of Korea 
(2019R1A2C2085965).
This research has made use of the KMTNet system operated by the Korea Astronomy and Space
Science Institute (KASI) at three host sites of CTIO in Chile, SAAO in South Africa, and 
SSO in Australia. Data transfer from the host site to KASI was supported by the Korea Research 
Environment Open NETwork (KREONET). 
This research was supported by the Korea Astronomy and Space Science Institute under the R\&D
program (Project No. 2023-1-832-03) supervised by the Ministry of Science and ICT.
The MOA project is supported by JSPS KAKENHI
Grant Number JSPS24253004, JSPS26247023, JSPS23340064, JSPS15H00781,
JP16H06287, and JP17H02871.
J.C.Y., I.G.S., and S.J.C. acknowledge support from NSF Grant No. AST-2108414. 
Y.S.  acknowledges support from BSF Grant No 2020740.
This research uses data obtained through the Telescope Access Program (TAP), which has been
funded by the TAP member institutes. W.Zang, H.Y., S.M., and W.Zhu acknowledge support by the
National Science Foundation of China (Grant No. 12133005). W.Zang acknowledges the support
from the Harvard-Smithsonian Center for Astrophysics through the CfA Fellowship. 
C.R. was supported by the Research fellowship of the Alexander von Humboldt Foundation.
\end{acknowledgements}

\end{document}